\pdfoutput=1 
\documentclass[a4paper]{article}

\usepackage{comment}

\usepackage{xurl}

\usepackage[colorlinks,allcolors=black]{hyperref}

\usepackage{pgfplots}
\pgfplotsset{compat=1.18}
\graphicspath{ {./figures/} }

\usepackage{subcaption}
\usepackage{calc}
\usepackage{amssymb}
\usepackage{amstext}
\usepackage{amsmath}
\usepackage{amsthm}
\usepackage{pslatex}
\usepackage{apalike}
\usepackage{algorithm2e}

\title{Implications of Edge Computing for Static Site Generation}
\author{Juho Vepsäläinen, Arto Hellas, and Petri Vuorimaa \and Aalto University}

\begin{document}
\maketitle

\abstract{Static site generation (SSG) is a common technique in the web development space to create performant websites that are easy to host. Numerous SSG tools exist, and the approach has been complemented by newer approaches, such as Jamstack, that extend its usability. Edge computing represents a new option to extend the usefulness of SSG further by allowing the creation of dynamic sites on top of a static backdrop, providing dynamic resources close to the user. In this paper, we explore the impact of the recent developments in the edge computing space and consider its implications for SSG.}



\vspace{10pt}

\textbf{The paper was accepted for WEBIST 2023 and the final version will be available in the conference proceedings. Note that this version has been edited to compile on arXiv and the final one is shorter due to a two-column layout.}

\section{\uppercase{Introduction}}
\label{sec:introduction}
Historically, websites have been hosted on servers serving static content~\cite{bernerslee1992}. The advent of content management systems (CMSs) brought about a dynamic approach that allowed editing the served contents with an online editor, leading to additional requirements and complexity from the server, including server-side rendering (SSR)~\cite{boiko2005,w3techs}. \textbf{Static site generators} (SSGs) that build static files out of dynamically edited contents were developed to mitigate the need for server-side rendering, yielding the possibility to serve the content with static file servers with little need for dynamic functionality. This possibility of serving static content -- coupled with an increased demand in throughput -- in part led to the emergence of \textbf{content delivery networks} (CDNs), which leverage a global network of servers for faster content delivery through geographical distribution.

With the emergence of commercial server providers and the decline of self-hosting, server farms were developed. On top of server farms, new methods for trading computational resources emerged, including the cloud computing market. Contemporary offerings allow paying based on the execution of individual function calls, potentially accounting for CPU and memory usage. This shift significantly contrasts the traditional trading of computational power, as the payment unit can be measured through individual computations rather than pieces of hosted hardware~\cite{lynn2017preliminary}. From the point of view of a computation resource vendor, this has enabled new economies of scale while encouraging custom hardware development.

The combination of these advancements -- CDNs and the more fine-grained control and billing of computation power -- has led to the emergence of \textbf{edge computing} as a viable option for web developers. While CDNs have benefits, edge computing allows programmability and selling function executions on top of CDNs. Edge computing has emerged as a viable option for software developers as it allows them to shape client requests and server responses at a scale near to the client, enabling faster response times~\cite{carvalho2021edge}. The shift to the edge has resulted in new technical solutions, such as edge-friendly databases, and the problem of cold starts familiar from cloud computing is becoming solved~\cite{eliminatingColdStarts}.



In the present study, we explore the impact of edge computing for static website hosting to evaluate how the ideas from static and dynamic realms may be mixed, answering the question \textit{What are the technical opportunities and challenges of edge computing for static website hosting?} A version of the question was previously posed in~\cite{vepsalainen2022bridging}, where the authors discussed the challenges of SSG when adjusting site contents and proposed an intermediate JSON representation format for site data. The work expands on a recent overview of edge computing research by \cite{cao2020overview} in the specific case of SSG.


The approach to answering the research question is two-fold. We first explore the advances of website rendering and hosting in Section~\ref{sec:background} to create a view of the recent movements in the space. Then, to illustrate the benefits of edge computing in practice, we explore the efficacy of rendering a blog platform using three rendering mechanisms and two popular edge providers, outlined in Sections \ref{sec:methodology} and \ref{sec:results}. The results of our experiment and the potential of edge computing for SSG are discussed in Section~\ref{sec:discussion}. Finally, Section~\ref{sec:conclusion} provides a conclusion and outlines directions for future study.

\section{\uppercase{Background}}
\label{sec:background}

In this section, we outline the main movements leading to the present state of creating websites and delivering websites. We also consider how these developments align with the emerging trend of edge computing in the web space.


\subsection{Evolution of Website Rendering Techniques}

Website rendering techniques have evolved since the beginning of the web to address the new requirements set for websites. The evolution has been supported by the growing market and the shift in use cases for the web as it grew from a site platform to an application platform as web applications became popular with the introduction of social networking and related trends. The growth of the web platform motivated the development of multiple website rendering techniques that each address specific pain points related to developing websites and web applications.

\subsubsection{Server Side Rendering}

Early websites developed in the 1990s were mainly static and served using static file servers. A static site consists of HTML pages, documents, and media, which can be read by the server from persistent storage and served to the client (usually a web browser) without further customization \cite{petersen2016}. Due to a need to provide a degree of interactivity and to allow changing the served data, dynamic functionality was added to the servers. Dynamic websites are typically stored in a format not directly renderable by the browser~\cite{petersen2016}. In the dynamic case, the server takes an incoming request, performs some actions in between, and generates a response that is then sent to the client. The process is commonly known as server-side rendering (SSR).


\subsubsection{Client Side Rendering and Single Page Applications}

SSR was the prevalent technology for building content for the web for over a decade until its slow decline in favor of client-side rendering (CSR) in the late 2000s and early 2010s. The move towards CSR stemmed from a potential for increased perceived usability as while SSR required the whole site to be reloaded per request, CSR allowed changing only the parts needed on a page using technologies such as JavaScript without forcing a refresh~\cite{flanagan1998}. A culmination point of this development was the emergence of single-page applications (SPA) in which it became possible to dynamically adjust the shown content based on the user interactions~\cite{mikowski2013single,ryan2021}.


\subsubsection{Static Site Generation}

Both SSR and CSR are complemented by static site generation (SSG). In SSG assets are compiled together to a format that can be hosted using a static file server \cite{newson2017} while coming with benefits related to security~\cite{petersen2016,camden2017}, fast page load times~\cite{petersen2016,camden2017}, scaling~\cite{petersen2016}, compatibility with versioning systems~\cite{camden2017}, and efficient resource usage~\cite{petersen2016}. 

Traditionally, SSGs have been a great fit for small content sites as in the worst case and the most naïve implementation, an SSG must recompile the entire site when the content changes. However, techniques such as incremental compilation enable an SSG to reuse the previous results while recompiling only the parts that a change made by the user affects.


There exists a wide variety of SSGs. For example, \url{https://jamstack.org/} enumerates over 350 SSGs (August 2023) in their listing \cite{jamstack2022} while \url{https://staticsitegenerators.net/} has over 460 SSGs (August 2023) \cite{ssg2022}.

\subsubsection{Jamstack}

Jamstack was introduced by Matt Biilmann at Smashing Conf in 2016 as a response to the weaknesses of the SSG model. It represents a change in thinking compared to the traditional web \cite{kumar2019} and shifts the perspective on how websites should be composed. The idea is to decouple content from the layout and then collect them together. The approach goes well with headless CMSs that expose their data through an API for third parties to consume~\cite{barker2017state}. Standard webhooks allow refreshing a website when the data changes~\cite{hoang2020jamstack}.

\begin{figure}
    \centering
    \includegraphics[width=10cm, bb=0 0 1243 750]{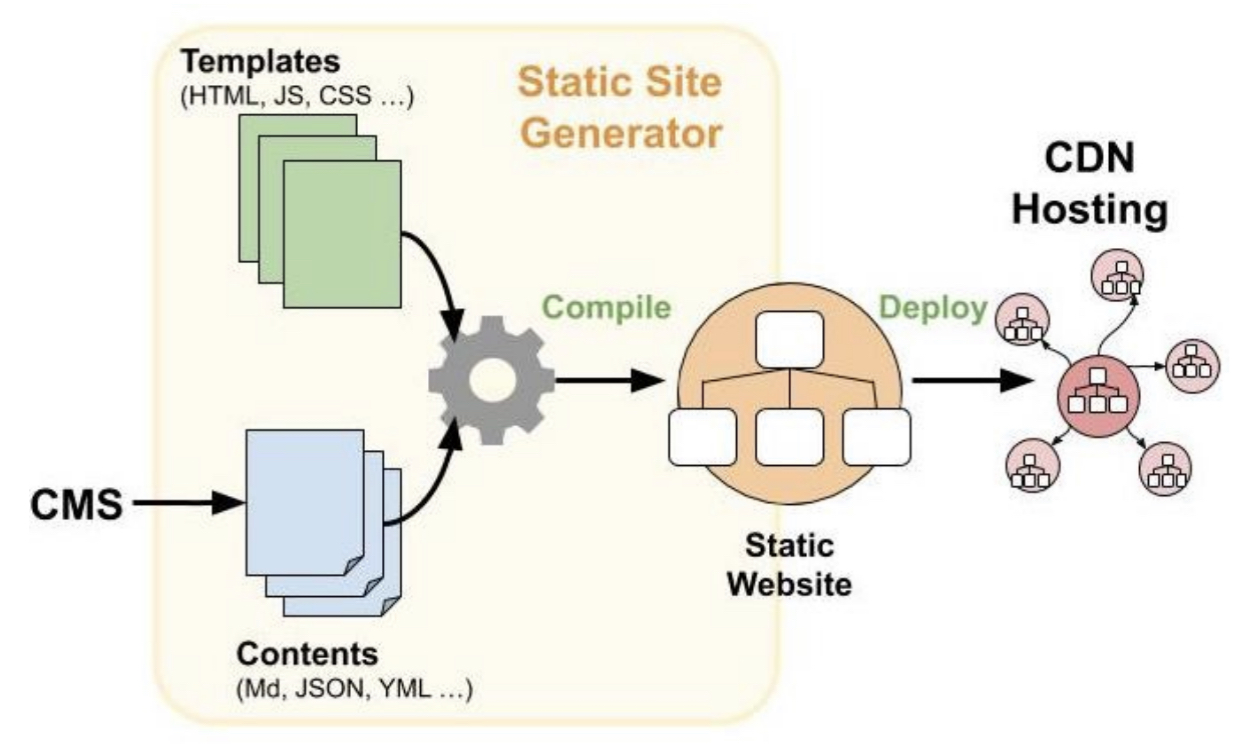}
    \caption{In Jamstack approach, data from a content management system is combined with templates (HTML, JS, CSS, etc.) that are then compiled using a SSG. The resulting static website is then deployed on a CDN hosting service. \cite{utomo2020building}}
    \label{fig:Jamstack}
\end{figure}

From a deployment point of view, Jamstack sites are still static and can be hosted through a static file server, therefore inheriting the SSG approach's benefits \cite{markovic2022could}. Jamstack relies on external services for dynamic functionality, such as authentication \cite{peltonen2021}. Due to their static nature, Jamstack sites can be hosted on CDNs and gain their benefits in terms of security and scalability, as with SSGs earlier. Figure \ref{fig:Jamstack} shows how the dynamic and static portions of Jamstack go together and how a Jamstack site is deployed on a CDN.

According to \cite{markovic2022could}, the hype around Jamstack is currently at its peak, and their findings indicate that although Jamstack is a promising approach, it may not become the de facto model for web development as there are concerns related to handling dynamic use cases and that is one of the main challenges the advocates of the Jamstack approach have to resolve in the coming years. Several early pain points of Jamstack have already been resolved through improved service offerings that cover features such as authentication or payment. The problem of previewing the impact of data changes has been alleviated to some extent through techniques such as incremental static regeneration \cite{markovic2022could}.

\subsubsection{Incremental Static Regeneration and Distributed Persistent Rendering}

Recent frameworks, such as Next.js, offer the possibility for SSR, CSR, and SSG, leading to hybrid functionality. Hybrid approaches enable developers to use the rendering technology that makes the most sense at a given time. On top of this, Next.js innovated a rendering method called incremental static regeneration (ISR), mixing SSG and SSR, that allows the use of SSG without rebuilding the entire site by shifting some of the work to on-demand~\cite{nguyen2022jamstack}. In the on-demand case where ISR is leveraged, pages are cached, and subsequent requests rely on the cache.

In 2021, Netlify introduced distributed persistent rendering (DPR). The idea of DPR is to address the shortcomings of ISR by providing atomic and immutable deploys consistent with the notion of Jamstack. In ISR, the users may see stale content on the first render by design, and this perceived shortcoming has been removed in DPR \cite{githubDistributedPersistent}.

\begin{figure*}
    \centering
    \includegraphics[height=5cm, bb=0 0 2241 236]{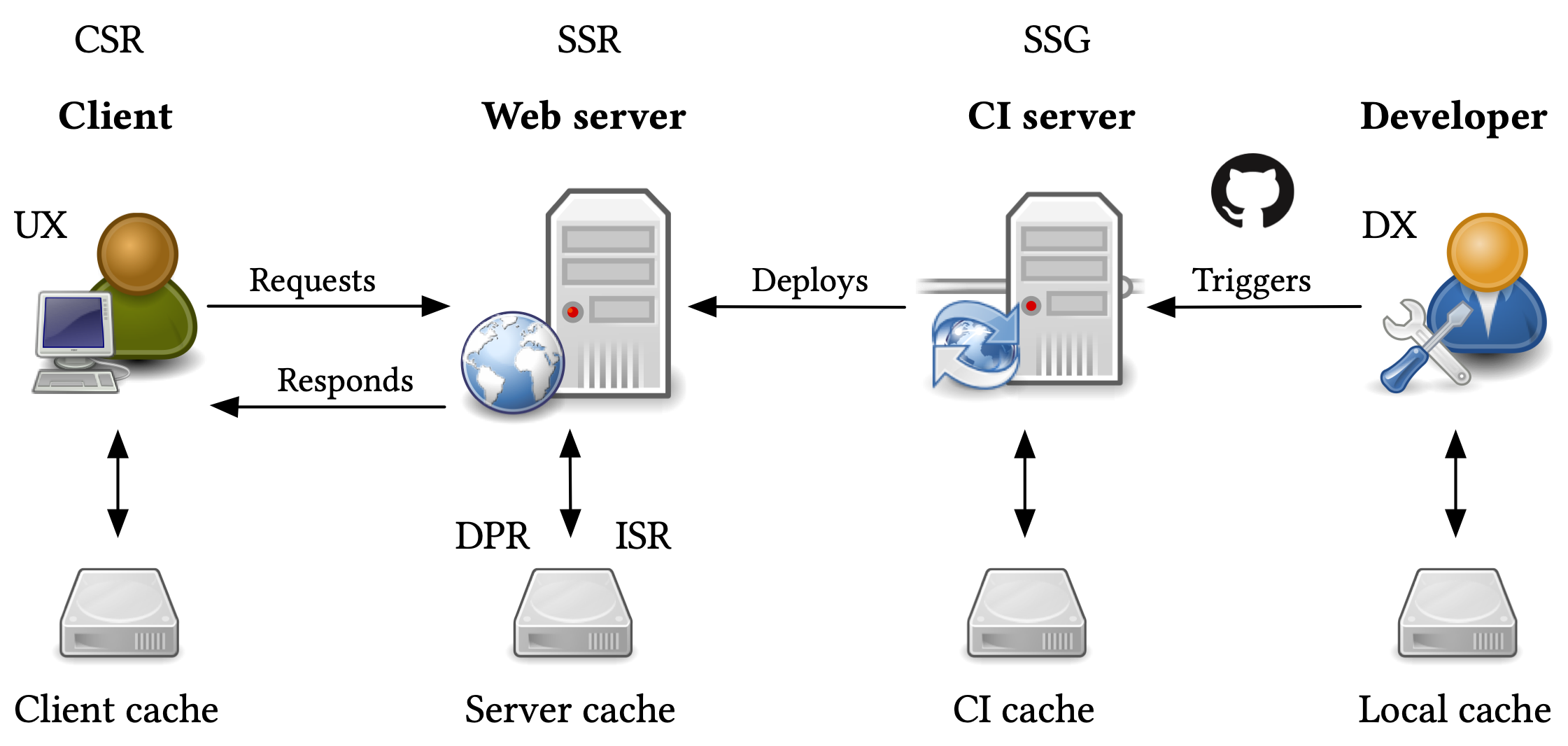}
    \caption{Workflow from a client to a developer. The workflow applies to traditional web and edge computing; the number of web servers can be scaled.}
    \label{fig:client-dev-ssg}
\end{figure*}

To understand how different rendering techniques relate to the client and the developer, Figure~\ref{fig:client-dev-ssg} summarizes them in a graphical form.

\subsubsection{Islands Architecture}

As discussed by \cite{vepsalainen2023rise}, islands architecture is a way to include dynamic portions to a static page and define strategies for loading them. Deferring loading allows pushing work performed by JavaScript to the future; some of the work may not occur depending on usage. As the architecture was formalized in 2019 \cite{jason2020}, there is not much experience in using it but at the same time solid adoption of \href{https://astro.build/}{Astro framework} leveraging the approach shows increasing developer interest.


\subsection{Evolution of Website Hosting}

Similar to website rendering techniques, the ways to host websites have evolved. The evolution of website hosting comes together with rendering techniques as they form a pair in the sense that hosting enables rendering at different levels of technical infrastructure, allowing new models for developing websites. 

\subsubsection{Rented Servers and Virtual Machines}

At the beginning of the web, companies and individuals had to maintain their servers. A whole hosting market emerged to make it easier for people to host their websites and applications. The early providers offered space for static websites and offered dedicated servers to rent. Later, virtual machines (VMs) emerged as an abstraction, decoupling hosting from hardware and enabling the sharing resources across multiple users. A key enabler here was HTTP/1.1, which provided the means to indicate the host to which a request was directed, in addition to the IP. 


\subsubsection{Content Delivery Networks}

With increasing demand and an acknowledgment that parts of the served contents were static and rarely changed, CDNs, such as Akamai, emerged~\cite{nygren2010akamai}. CDNs provided both the possibility of distributing requests over a broader range of servers to decrease individual server load and to respond to requests from servers close to the client, thereby reducing the latency experienced by the user~\cite{triukose2011measuring}.

\subsubsection{Cloud and Serverless Computing}

Cloud computing was a movement in offering computing resources that abstracted away physical hardware. One could still buy a virtual machine when buying resources from cloud computing providers. Still, the location of the virtual machine might have been unclear, and it was also possible that the physical machine running the virtual machine could change dynamically. The infrastructure built to support cloud computing slowly led to the emergence of the serverless computing paradigm, where the notion of starting a server was abstracted away, and developers instead defined entry points to applications. In serverless computing, functions are triggered on demand while having access to databases~\cite{jonas2019}.


\subsubsection{Edge Computing}

Edge computing represents the next step in how and where computation occurs. Edge computing is a natural evolution over the CDN approach as instead of only serving resources; it enables computation close to the client on demand~\cite{weisong2016}. The distributed approach leads to new technical challenges as traditional ways of thinking about aspects, such as databases, must be reconsidered to be compatible with a global infrastructure. In general, edge computing shows promise in improving web page and content rendering performance~\cite{zhu2013improving,viitanen2018low}, reinvigorating discussions on making informed decisions on what content to serve to account for network quality~\cite{zhu2013improving}.

\subsubsection{Discussion}

The latest developments in rendering techniques and edge computing allow us to address the traditional limitations of SSG and Jamstack while gaining their benefits. Most importantly, edge computing provides a way to intercept user requests before they reach the file server. Alternatively, the edge network can work as a server and return suitable payloads to the client directly. Perhaps more interestingly, edge computing enables the development of hybrid websites where some portions are static and others are dynamic. The islands architecture is a good example of an approach ready to leverage edge computing.



There are some concerns related to the lock-in potential of edge platforms. At the same time, initiatives such as \href{https://wintercg.org/}{WinterCG} provide hope of collaboration to make JavaScript-based edge runtimes compatible with each other. In the ideal case, developers should be able to move edge workers from one platform to another with minimal effort. 

\section{\uppercase{Methodology}}
\label{sec:methodology}
To illustrate the implications of edge computing for SSG, we benchmark a statically hosted site against one served from an edge platform. We hypothesize that their performance is close to each other, although we expect the latter solution to come with a slight performance cost depending on the use of caching. To provide a third point of view, we examine the impact of ISR as it is a technique between SSG and SSR.

\subsection{Platform and implementation}

For the present study, we explored the efficacy of a blog platform with the following constraints\footnote{For replication and analysis of the implementation, the source code for the project has been made available at \url{https://github.com/bebraw/ssg-benchmark}}: 

\begin{enumerate}
    \item There are three variants to compare: static site generation (SSG), pure edge server-side rendering (SSR), and edge server-side rendering with ISR, which leverages Cloudflare KV for caching \footnote{We have adapted Matteo Rigon's implementation for this purpose, and the original version can be found at \url{https://reego.dev/blog/achieving-isr-on-cloudflare-workers}.}
    \item All variants are implemented using TypeScript.
    \item The static variant is generated using an ad hoc implementation based on ES2015 templates for templating. The edge variants use the same logic.
    \item The static variant is hosted on both \href{https://pages.cloudflare.com/}{Cloudflare Pages} and \href{https://www.netlify.com/}{Netlify} so it will be measured twice to see the impact of the platform.
    \item The edge variants are implemented using \href{https://developers.cloudflare.com/workers/}{Cloudflare workers}.
    \item The site to test mimics a blog with a blog index and individual pages.
    \item Styling and images are kept out of scope to keep the test case simple and to avoid loading costs.
    \item All variants fetch content from a small server, returning pseudorandom data for repeatability.
    \item Each implementation had a fixed 100ms delay to simulate the cost of server-side logic.
\end{enumerate}

Cloudflare and Netlify platforms were chosen; both offer edge computing facilities. Cloudflare is a company that started as a CDN provider and has then expanded to hosting and edge computing, which are natural extensions to the CDN business. Cloudflare has developed solutions to cloud computing problems, including approaches for eliminating cold starts related to starting edge workers \cite{eliminatingColdStarts}. Netlify, similar to Cloudflare, provides edge computing capabilities and a Git-connected way to deploy applications on their platform \cite{netlifyhosting}. For the scope of the present work, Netlify is used only as a static host. These platforms were chosen by their relative popularity in the developer community, and an expanded study should include more options to benchmark.


\subsection{Measurement of performance}

For performance measurements, we used \href{https://playwright.dev/}{Playwright} with \href{https://developer.chrome.com/docs/lighthouse/}{Google Lighthouse}. We created a test suite that is run against the blog site variants, intended to capture any differences in performance. For the present study, each blog site variant hosted a hundred blog posts, and when measuring performance, we focused on First Contentful Paint and Server Response Time. In addition, we used \href{https://www.npmjs.com/package/autocannon}{Autocannon} to capture rough throughput as responses per second and latency for each variant. Lighthouse and Autocannon have commonly used tools for assessing website performance, and blogs are a common archetype in web applications.

Following \cite{herivcko2021towards}, who noted that performing Lighthouse performance audits five times reduces variability in results significantly in a reasonable time, we executed the tests five times. This is in line with the Lighthouse documentation that suggests that measuring the median of five runs is twice as stable as measuring a single run~\cite{lighthousevariability}.

For the Lighthouse tests, we measured the rendering performance of the blog index page (listing 100 blog links) and the performance of a blog page (showing a blog entry), and we throttled the network using mobile (1.6 Mbps down / 750 Kbps up) with 150 ms latency. For the Autocannon tests, we measured the performance of the blog index page. We wrote the test to run for 30 seconds per variant to decrease the impact of variability in connection quality. Before every ISR variant-related test, the cache was emptied manually to avoid skewing results.

\subsection{Threats to validity}

The tests we perform are black-box by their nature. In other words, we do not control and know anything about the underlying infrastructure. There may be significant differences at the infrastructure level and technical implementations we are unaware of. However, the platforms we benchmark claim to implement the edge paradigm and expose related APIs.

Another threat to validity has to do with the scope of testing. Given we test from a single location, we do not test the scalability of the approach from a global perspective. Global scaling is considered one of the selling points of the CDN approach, but it is out of the scope of the study.

Our test project is synthetic and reflects only a simple static use case. In practice, web applications can be far more complex and dynamic by nature. The test project provides a baseline for more dynamic tests that build on top of static functionality. 

\section{\uppercase{Results}}
\label{sec:results}
In the following subsections, we show Lighthouse and Autocannon results separately.

\subsection{Lighthouse results}

Lighthouse scores pages from zero to a hundred based on the categories: performance, accessibility, best practices, SEO, and PWA. While we focused on First Contentful Paint and Server Response Time, we also briefly studied the other Lighthouse metrics. For each page tested, the performance, accessibility, and best practices metrics received a full score of hundred. SEO varied between 82 and 91, suggesting that the implementation was missing a meta description and the blog page implementation had too tiny tap targets on mobile.

For each variant, the First Contentful Paint (FCP) and Server Response Time (SRT) values have been listed in Table~\ref{table:measurement-results}. Time to Interactive (TTI) followed FCP closely in this scenario. The values have been rounded to the closest value and are provided in milliseconds (ms). The first test run and the subsequent four test runs are reported separately in the table. 

\begin{table*}[ht]
 \vspace{-0.2cm}
 \caption{Summarized measurement results (each result is given in ms). CF = Cloudflare, FCP = First Contentful Paint, SRT = Server Response Time. The suffix index indicates the performance of the index page with 100 blog post links, while the suffix post indicates the performance of an individual blog page with the blog post contents.}
 \begin{tabular}{ |l|c|c|c|c|c|c| }
 \hline
    & \multicolumn{3}{c|}{FCP} & \multicolumn{3}{c|}{SRT}\\
 \hline
 Run & 1 & 2-5 (med.) & 2-5 (avg.) & 1 & 2-5 (med.) & 2-5 (avg.)\\
 \hline
CF SSR index & 1053 & 1028 & 1039 & 283 & 282 & 276 \\
CF SSR post & 1030 & 991 & 1053 & 280 & 263 & 322 \\
CF ISR index & 895 & 879 & 889 & 145 & 131 & 135 \\
CF ISR post & 879 & 880 & 876 & 166 & 159 & 148 \\
CF SSG index & 919 & 1026 & 987 & 160 & 272 & 227 \\
CF SSG post & 873 & 860 & 862 & 145 & 128 & 133 \\
Netlify SSG index & 963 & 880 & 924 & 241 & 140 & 186 \\
Netlify SSG post & 955 & 861 & 872 & 241 & 149 & 170 \\
 \hline
 \end{tabular}
 \label{table:measurement-results}
\end{table*}

\subsection{Autocannon results}

\begin{figure}
\centering
\begin{tikzpicture}
 \begin{axis}[ymode=log, width=.96\textwidth, legend pos=north west, xlabel=$Percentile$, ylabel=$Time (ms)$]
  \addplot table [x=a, y=b, col sep=comma] {autocannon.csv};
  \addplot table [x=a, y=c, col sep=comma] {autocannon.csv};
  \addplot table [x=a, y=d, col sep=comma] {autocannon.csv};
  \addplot table [x=a, y=e, col sep=comma] {autocannon.csv};

  \addlegendentry{Cloudflare SSR}
  \addlegendentry{Cloudflare ISR}
  \addlegendentry{Cloudflare SSG}
  \addlegendentry{Netlify}
 \end{axis}
\end{tikzpicture}
  \caption{Autocannon latency per percentile over each variant over a thirty-second interval shown using a logarithmic scale. Note the peak at the end. Also, note that ISR and SSG follow each other as the cost of ISR is visible only on the first render, and due to the number of runs it vanishes.}
  \label{graph:autocannon-latencies}
\end{figure}
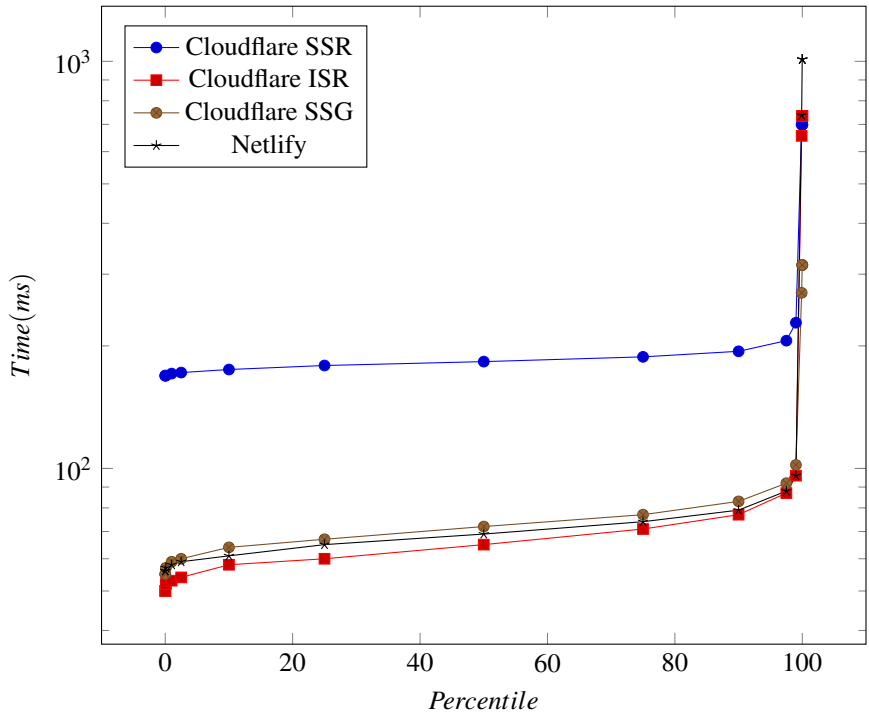

For measuring the application's throughput, we utilized Autocannon, studying how the latency behaves over the requests in the 30-second time, focusing on the blog index page for each variant. Figure~\ref{graph:autocannon-latencies} outlines the latency per percentile, which shows sub-100 millisecond latencies for most requests. In the Figure, the 100 ms latency embedded in the blog code to highlight additional server-side logic is visible in the SSR option, as the option does not benefit from caching. The differences would be negligible if we omit the additional 100 ms latency.

In general, the Autocannon results are somewhat consistent with the Lighthouse server response times, although the Lighthouse server response times show more variance, perhaps due to the fewer tests. In the Autocannon test, we view that the 100\% percentile could be safely dropped as it represents individual outliers -- on average, over the thirty seconds, the Autocannon tests yielded between 18,000 and 30,000 responses, which our single-computer test setup may partially limit. 

\section{\uppercase{Discussion}}
\label{sec:discussion}

Given the measurements, we can see that the latencies of edge platforms are low, especially for the SSG and ISR cases. SSR is expected to come at a cost as there is more processing. The difference became apparent due to the artificial delay added to SSR, and in practice, the delay could be even more visible due to database requests and further work to perform per request. The benefit of ISR is that it allows us to avoid build time work and shift it to runtime at the cost of potentially stale cache for the client.


It would be possible to discard the entire ISR cache during deployment to address the staleness issue. Doing this would shift the implementation closer to Netlify's Distributed Persistent Rendering (DPR)~\cite{netlifyDistributedPersistent}, which seeks to address the shortcomings of ISR by providing atomic and immutable deploys consistent with the idea of Jamstack. Furthermore, assuming the ISR cache has a staleness factor, such as time, related to it, the Stale While Revalidate (SWR) technique could be applied to return stale results while generating a new page in the background. In this case, the next request would yield fresh results~\cite{reegoIncrementalStatic}.


\subsection{When to apply ISR?}

Given there's a cost related to SSG and especially to building sites on content change, the question is when it becomes beneficial to apply techniques such as ISR. There is added complexity for small sites due to having to use a framework or program on the edge. For highly dynamic use cases where the content changes often, the added complexity may be worth it, as otherwise, you would have to build the site constantly. For example, for social media platforms with rapidly changing content, static site generation might not be a feasible option -- in such a case, one could rely on hybrid rendering approaches, which were scoped out from the present work. It could be argued that techniques, such as incremental compilation, can significantly decrease the cost of doing this.

\subsection{No cold start cost at Cloudflare}

Interestingly, Cloudflare does not seem to have a cost associated with a cold start, while Netlify has a cold start penalty, as evidenced in the server response time measurements. The lack of penalty is a good sign, which means response times are more predictable for developers. At the same time, we could observe, however, that an individual response might occasionally take up to a second at the extreme outliers while, generally, response speeds were stable.

Our measurement server latencies (SR) generally seem low and are within the 300 ms range. The rest of the cost occurs on the browser side (FCP), implying that development practices matter as developers can optimize this cost. It is also good news for framework authors, given they can use the findings to optimize asset delivery.



\subsection{Shift of JavaScript frameworks towards the edge}

The latest generation of JavaScript frameworks, such as Astro or Qwik, are compatible with the edge out of the box and support the most popular edge platforms as deployment targets while coming with static functionality as well. They support hybrid rendering and allow developers to choose what technique to use and where. The results of the study support this movement as there are clear benefits to SSG combined with edge computing.

\subsection{Potential of edge-powered islands}

Since the edge provides simple ways to encapsulate logic within workers, developers can leverage islands architecture on top of their static sites. Using an appropriate strategy, the idea is to encapsulate dynamic functionality behind an edge worker and call that within an island. To simplify the task, \href{https://github.com/11ty/is-land}{11ty/is-land} implements multiple strategies while allowing any framework to be used for rendering the islands, making it a good companion for the edge. The idea would be to leverage the \textit{template} element of HTML while pointing to the edge endpoint that implements the island contents.

\href{https://www.11ty.dev/}{Eleventy}, a popular SSG, implements edge support natively through shortcodes included in templates \cite{11tyEdge}. The feature is experimental and works only with the Netlify Edge platform \cite{11tyEdge}. First-class support for the edge on an SSG tells about the direction and the fact that tool authors have recognized the potential of the edge. The same is visible in solutions like Astro that allow hosting and processing on edge while supporting pure SSG. 

Cloudflare research team devised the fragments architecture, and the target of this work was to allow building micro-frontends using Cloudflare Workers\cite{cfFragments}. The idea is consistent with edge-powered islands and approaches it from a vendor point of view while considering legacy and mixed systems enabled by micro-frontends where teams can develop using technologies they prefer. Cloudflare researchers' work implies a crossing point between micro-frontends, islands, and edge computing, which alone may be a direction worth exploring in further study as a technological intersection.

Edge-powered islands come with challenges related to a state shared by multiple islands. It is also likely more suitable for cases with limited interactivity than experiences where the whole page has to be dynamic by definition. In other words, edge-powered islands expand the types of applications that can be developed on top of SSG but encounter limits in highly dynamic use cases. 

\section{\uppercase{Conclusion}}
\label{sec:conclusion}

We started this paper by asking the question \textit{What are the technical opportunities and challenges of edge computing for static website hosting?} and found out the intersection expands the usefulness of SSG by allowing more dynamic use cases to be covered on top of it. There are clear opportunities in leveraging architectures like islands architecture on top of a static site. The performance of edge platforms seems reasonable enough in terms of latency, and techniques, like ISR, address problems related to SSG build speed. 



Our empirical evaluation demonstrated how SSG and edge computing can work together to enable performant websites and applications to be developed, in part yielding evidence on the efficacy of mixing web technologies as asked for in~\cite{vepsalainen2022bridging}. That said, there are still open questions related to techniques, their applicability in other environments, and their limitations. Furthermore, there are questions related to the costs of the platforms in comparison to the cloud and self-hosting. It's undeniable developing a comparable infrastructure yourself would be cost-prohibitive for many but at the same time not all applications require the same capabilities.


On top of build and server infrastructure, there are layers of techniques related to leveraging caching, prefetching, and pushing work to the client. These techniques are often orthogonal and may be used to complement server-side optimizations. In terms of research, it would be valuable to understand which optimizations can be done at each level how much can they contribute towards the overall performance of a web service, and at what cost.

There are also questions related to reproducing the study results globally. Given edge infrastructure operates on top of CDN, the assumption is that the results should be fairly consistent across the globe depending on CDN density. That starkly contrasts traditional architecture where the server is in a specific location. Measuring the difference and reproducing the study with a global scope would be worthwhile. To help with this goal, our implementation and evaluation code are available on GitHub at \url{https://github.com/bebraw/ssg-benchmark}. 

\bibliographystyle{apalike}
{\small
\bibliography{references}}

\end{document}